\documentstyle[12pt]{article}
\title{
Thermal fission rate around super-normal phase
transition\\}
\author{K. Hagino,$^1$ N. Takigawa,$^1$ and M.Abe$^2$\\ \\
{\it $^1$ Department of Physics, Tohoku University, 980-77 Sendai, Japan}\\
{\it $^2$ Faculty of Science and Engineering,}\\
{\it Ishinomaki Senshu University, 986 Ishinomaki,Japan}}
\date{}

\begin{document}
\baselineskip=9mm
\maketitle

\begin{center}
{\bf Abstract}
\end{center}

Using Langer's $Im F$ method, we discuss the temperature dependence
of nuclear fission width in the
presence of dissipative environments.
We introduce a low cut-off frequency to the spectral density of
the environmental oscillators in order to mimic the pairing gap.
It is shown that the decay width rapidly decreases at
the critical temperature, where the phase transition from super
to normal fluids takes place.
Relation to the recently observed threshold for the dissipative
fission is discussed.

\bigskip

\noindent
PACS number(s):
24.75.+i, 
25.70.Jj, 
24.60.Dr, 
05.40.+j 

\newpage

Fission of a hot nucleus has attracted much interests of
nuclear physicists in the past several years
to study nuclear dissipation together with deeply inelastic
heavy-ion scattering[1--3].
It is known that statistical codes to calculate the dacay of a
compond nucleus significantly underestimate the experimentally observed
prefission neutron, charged particle, and $\gamma$-ray
multiplicities at high excitation energies if
the original Bohr-Wheeler formula for the fission width is used,
though it works pretty well at low energies[4--7].
One way to understand this fact is to consider that fission is hinderd
by nuclear dissipation.
Based on this idea, Thoennessen and Bertsch have analysed
fission data on prefission neutron, charged particle and
$\gamma$ ray multipicities for various systems by using statistical codes,
and obtained systematics of the threshold for dissipative fission
\cite{TB93,V94}.
This systematics has been confirmed
by experimentally studying the excitation energy dependence of the fission
probability in $^{200}$Pb compound nuclei\cite{F94}.

On the other hand, the nuclear dissipation does not
play any significant role in spontaneous fission
because of the strong pairing correlation between nucleons\cite{DW87,BBBV90}.
When one discusses nuclear fission at moderate excitation
energies, one has to take into account the
temperature dependence of the pairing gap.
The pairing gap decreases with temperature and
the nucleus eventually undergoes a phase transition from a superfluid
to a normal fluids[10--13].
It is the purpose of this paper to investigate
the effects of the super-normal phase transition on the
fission width at finite temperatures.
To this end, we use Langer's $Im F$ method,
where the decay width of a metastable state is related to
the imaginary part of the free energy[14--18].
In this method one can describe the decay process for a very wide
range of temperature, i.e. from zero temperature, where
the decay process is governed by the quantum tunneling,
to high temperatures, where thermal decay dominates\cite{HTB90}.
Also, the method can be applied to
a system with many degrees of freedom\cite{GOW87}.

We consider a system where a macroscopic degree of freedom $q$
is coupled to environmental heat bath.
In the problem of fission, $q$ corresponds to
the fission coordinte.
We assume the following Lagrangian for this system\cite{CL81}
\begin{equation}
L=\frac{1}{2}M(q)\dot{q}^2-V(q)+\sum_i\frac{1}{2}m_i(\dot{x}^2_i-\omega_i^2
x^2_i)-\sum_ic_ix_if(q)+\sum_i\frac{c_i^2f(q)^2}{2m_i\omega_i^2}
\end{equation}
where $\{x_i\}$ and $\{\omega_i\}$ are the coordinates of the environmental
oscillators and the corresponding excitation energies, respectively.
$V(q)$ is a potential for the macroscopic degree of
freedom, which has a local minimum and a maximum at
$q=q_0$ and $q=q_b$, respectively.
$M(q)$ and $f(q)$ are the mass of the macroscopic motion and
the coupling form factor, respectively. We assume general functions
of $q$ for them\cite{TA90,BZW94}.
The last term is the so called counter term which cancels the static
potential renormalization due to the coupling between
the macroscopic and the environmental degrees of freedom \cite{CL81}.
Takigawa and Abe have suggested
that, in contrast to heavy-ion fusion reactions at
subbarrier energies where the static potential renormalization
plays an important role in enhancing the fusion cross section
over the predictions of a one dimensional potential model\cite{THAB94,THA95},
the static potential renormalization in the fission
problem can lead to two opposite effects, i.e. it could
either lower or increase the effective fission barrier compared with the
bare potential barrier, thus leading to either hindrance
or enhancement of the fission rate, depending on the properties of the
coupling form factor $f(q)$ \cite{TA90}.
Both cases lead to a temperature dependent
fission barrier hight\cite{BSAA93,V94}.
In this paper, however, we introduce the counter term because
we do not know yet much about the coupling form factor.

In order to obtain the free energy, we first express the partition
function in
the path integral form. After integrating out the environmental degrees
of freedom,
the partition function at the temperature $k_BT=1/\beta$
takes the form\cite{GSI88}
\begin{equation}
Z(\beta)=\int{\cal D}[q(\tau)]e^{-S_{eff}[q(\tau)]/\hbar},
\end{equation}
where the path integral is performed over all the periodic paths
with the period $\beta\hbar$. The effective Euclidean action
$S_{eff}[q(\tau)]$ is given by
\begin{equation}
S_{eff}[q(\tau)]=\int^{\beta\hbar}_0d\tau \left(\frac{1}{2}M(q(\tau))
\dot{q}^2+V(q(\tau))\right)
+\frac{1}{2}\int^{\beta\hbar}_0d\tau\int^{\beta\hbar}_0d\tau'
k(\tau-\tau')f(q(\tau))f(q(\tau'))
\end{equation}
with the influence kernel $k(\tau)$ \cite{TA90,GSI88}
\begin{equation}
k(\tau)=\sum_i\left[\frac{c_i^2}{m_i\omega_i^2}:\delta(\tau):
-\frac{c_i^2}{2m_i\omega_i}
\frac{\cosh[\omega_i(|\tau|-\frac{1}{2}\beta\hbar)]}
{\sinh(\frac{1}{2}\hbar\omega_i\beta)}\right]
\end{equation}
where
\begin{equation}
:\delta(\tau):=\sum^{\infty}_{n=-\infty}\delta(\tau-n\beta\hbar)
\end{equation}
is a generalized delta function with period
$\beta\hbar$.

We consider now a high temperature regime, where
the decay of a metastable state is governed by the thermal hopping.
Evaluating the path integral in eq.(2) in the saddle point approximation
and using the relation between the decay width $\Gamma$ and the
imaginary part of the free energy\cite{L67}
\begin{equation}
\Gamma(T)=-\frac{2}{\hbar}\frac{T_c}{T}Im F,
\end{equation}
$T_c$ being the cross over temperature where the transition
between the thermal activated decay and the quantum tunneling
occurs,
we find that the decay width at temperature $T$ can be
expressed as\cite{TA90}
\begin{equation}
\Gamma=\frac{\omega_0}{2\pi}\frac{\omega_R}{\omega_b}
\sqrt{\frac{M(q_0)}{M(q_b)}}f_qe^{-\beta V_b},
\end{equation}
where $\omega_R$ is defined as $2\pi k_BT_c/\hbar$.
$\omega_0,\omega_b$, and $V_b$ are
the curveture of the potential barrier $V(q)$ at the local minimum $q=q_0$,
that at the barrier position $q_b$, and
the height of the potential barrier i.e., $V_b=V(q_b)-V(q_0)$, respectively.
$f_q$ is the quantum correction factor due to the
quantum fluctuation of the paths around the classical
paths $q_{c\ell}(\tau)=q_b$, $q_{c\ell}(\tau)=q_0$,  and is given by
\begin{equation}
f_q=\prod^{\infty}_{n=1}
\frac{\nu_n^2+\omega_0^2+\left(\frac{df}{dq}\right)_{q=q_0}^2
\nu_n\hat{\gamma}_0(\nu_n)}
{\nu_n^2-\omega_b^2+\left(\frac{df}{dq}\right)_{q=q_b}^2
\nu_n\hat{\gamma}_b(\nu_n)}
\end{equation}
where $\nu_n=2\pi n/\beta\hbar$ is the Matsubara frequencies.
$\hat{\gamma}$ is the Laplace transform of the retarded friction
kernel \cite{TA90},
and is given by
\begin{equation}
\hat{\gamma}(z)=\frac{1}{M(q)}\sum_i\frac{c_i^2}{m_i\omega_i^2}
\frac{z}{z^2+\omega_i^2}
\end{equation}
The subscripts 0 and $b$ in eq.(8) denote that the quantities with
those indices should be evaluated at $q=q_0$ and $q=q_b$, respectively.
The crossover temperature $T_c$ is identified with the highest
temperature at which the quantum correction factor $f_q$ diverges
\cite{GOW87}.
In the absence of environments, this prescription assigns $k_BT_c$ to be
$\hbar\omega_b/2\pi$. This is consistent with the
earlier observation by Affleck on the crossover temperature \cite{L67}.
It should be noticed that Langer's $Im F$ method implicitly assumes that
the coupling of the macroscopic degree of freedom to the environmental
degrees of freedom is strong enough to assure that the
system is always in a thermal equillibrium.

We now apply eq. (7) to the problem of the fission of a hot nucleus.
Following ref.\cite{DW87} we introduce a low cutoff frequency
$\omega_c$ to the distribution of the
environmental oscillators in order to mimic that
there is no nuclear levels below the two quasi particle
state in even-even nuclei.
Accordingly, we set the cutoff frequency to $2\Delta(T)/\hbar$,
$\Delta(T)$ being the pairing gap at the temperature $T$,
and take the spectrum density of the
environmental oscillators as\cite{DW87}
\begin{equation}
J(\omega)\equiv
\frac{\pi}{2}\sum_i\frac{c_i^2}{m_i\omega_i}\delta(\omega-\omega_i)
=\eta(\omega-\omega_c)\theta(\omega-\omega_c)
\end{equation}
where $\eta$ is the friction constant\cite{CL81}.
Note that $\omega_c=\infty$ and $\omega_c=0$ correspond
to two extreme cases where there is no dissipation at all and where
the spectrum density is given by the usual Ohmic dissipation, respectively.
The former and the latter cases give the Bohr-Wheeler formula and the
well known Kramers's formula at moderate to strong
friction for the decay rate,
respectively, with a quantum correction factor\cite{GOW87}.
The temperature dependence of the cutoff frequency $\omega_c$ is given
by solving the thermal gap equation[10--13]
%
\begin{equation}
\Delta=\frac{G}{2}\sum_k\frac{\Delta}{\sqrt{(\epsilon_k-\lambda)^2+\Delta^2}}
\tanh\left(\frac{\sqrt{(\epsilon_k-\lambda)^2+\Delta^2}}{2k_BT}\right)
\end{equation}
where $G$ and $\lambda$ is the strength of the pairing interaction and
the chemical potential, respectively.
For the spectrum density given by eq.(10), eq.(9) for the Laplace
transform of the damping kernel reads
\begin{equation}
\hat{\gamma}(z)=\frac{\eta}{M(q)}
+\frac{2}{\pi}\frac{\eta}{M(q)}\left(\frac{\omega_c}{z}
\log\frac{\omega_c}{\sqrt{\omega_c^2+z^2}}-\tan^{-1}\frac{\omega_c}{z}
\right)
\end{equation}
Note that the second term in this equation vanishes when the cutoff
frequency $\omega_c$ is set zero.

We now apply the above arguments to the fission of $^{248}$Cf.
We take the reduced mass for the symmetric fission for $M(q)$ and
the potential given in ref.\cite{GJW83} for $V(q)$.
$\hbar\omega_0, \hbar\omega_b, q_b$ and
$V_b$ then take the values of
1.18 MeV, 1.06 MeV, 3.4 fm, and 3.67 MeV, respectively.
We take the value 20$\times$10$^{21}$/sec.\cite{BSAA93}
for the reduced dissipation coefficient $\beta\equiv \eta/M$,
and assume a bi-linear coupling form factor i.e., $f(q)=q$.
Since we are interested in the effects of pairing in the
super to normal transition region,
we use a simplified expression for the temperatre dependent pairing gap,
\begin{eqnarray}
\Delta(T)&=&k_BT_c^{pair}\sqrt{\frac{8\pi^2}{7\zeta(3)}(1-(T/T_c^{pair})}
{}~~~~~~~(for~T<T_c^{pair}) \\
&=&0~~~~~~~~~~~~~~~~~~~~~~~~~~~~~~~(for~T>T_c^{pair})
\end{eqnarray}
which is valid near the transition temperature\cite{IG79}.
In eq.(13) $\zeta$ is the zeta function and  $T_c^{pair}$
the critical temperature for the super-normal phase transition.
We assign the pairing gap at zero temperature
to be 12/$\sqrt{A}$, $A$ being the
mass number of a nucleus, and estimate
the critical temperature $T_c^{pair}$ using
the relation $T_c^{pair}\sim 0.567\Delta_0$\cite{CDP83,IG79}.

Figure 1 shows the crossover temperature $T_c$
as a function of the cutoff parameter $\hbar\omega_c$.
This is given by solving
\begin{equation}
\omega_R^2+\omega_R\frac{\eta}{M}-\omega_b^2+\frac{\eta}{M}
+\frac{2}{\pi}\frac{\eta}{M}\left(\frac{\omega_c}{\omega_R}
\log\frac{\omega_c}{\sqrt{\omega_c^2+\omega_R^2}}
-\tan^{-1}\frac{\omega_c}{\omega_R}\right)=0
\end{equation}
It should be remarked that in calculating the
decay rate based on eq.(7) the crossover temperature
$\omega_R$ has to be evaluated at each temperature $T$ with
corresponding cuttoff frequency $\omega_c$ i.e., one must solve
eq.(15) by treating $\omega_c$ as though it is independent of temperature.
Otherwise, one cannot recover the decay rate formula of Kramers modified
by the quantum correction factor at temperatures higher than $T_c^{pair}$,
where the pairing gap vanishes.
The solid line in fig.1 is the solution of eq.(15).
The dashed line is the crossover temperature in the absence of environments,
i.e. $\hbar\omega_b/2\pi$.
If one sets $\omega_c$ to be zero, the crossover temperature is
given by $(\sqrt{1+\alpha^2}-\alpha)\hbar\omega_b/2\pi$,
$\alpha$ being $\eta/2M\omega_b$ \cite{GOW87}. This value is denoted
by the dotted line in the figure.
The crossover temperature gradually decreases as the cut-off frequency
decreases reflecting the increasing dissipation \cite{DW87}.

Figure 2 shows the quantum correction factor given by eq.(8)
as a function of the temperature.
In the limits of $\omega_c\to 0$ and $\infty$, the infinite
product in eq.(8) can be simplified by using $\Gamma$
function\cite{GOW87,FT92}.
In the case of finite $\omega_c$, one has to evaluate it
directly until one gets convergence. In general cases, however,
this is a fairly difficult numerical task because the ratio for
each $n$ in eq.(8) never becomes sufficiently close to one
even for very large $n$. Consequently, numerical errors
accumulate as one performs the production many times.
In our applications, where we used a constant mass and a bilinear coupling,
the infinite product series converged.
The dashed and the dotted lines are the quantum correction factor
in the limit of  $\omega_c\to 0$ and $\infty$, respectively.
The solid line is the quantum correction factor when
the lower cutoff for each temperature has been introduced.
The left and the right arrows in the figure show the crossover temperature
from a quantal to a thermal decay, i.e. $T_c$=0.169 MeV, in the absence
of environment and the transition temperature from super to normal
fluids, i.e. $T_c^{pair}$=0.432 MeV.
The solid line coincides with the dotted line at temperatures
higher than $T_c^{pair}$, as is expected.
Note that the quantum correction factor approaches to one at
high temperatures.

The decay rate for this system is shown in figure 3 as a function
of the temperature.
The meaning of each line is the same as that in fig.2.
We observe a sudden decrease of the decay rate at
the critical temperaure $T_c^{pair}$.
This behaviour agrees with that found in ref.\cite{YAF93},
where the diffusion of a heavy particle in metal was studied
by taking a superconducting phase transition of the environmental
electrons into account. Notice that the cusp behaiour in the transitional
region will be smeared out to some extent in actual cases, for example,
by the gradual disappearance of the pairing gap with temperature.

In summary, we made use of the $Im F$ method of Langer
to discuss the fission dynamics of
hot nuclei in the presence of a dissipative environment.
We modified the Caldeira-Leggett model by introducing a low cutoff frequency
in order to mimic the effects of nuclear superfluidity due to pairing
interaction. We took into account
the temperature dependence of the pairing gap,
and thus the phase transition from a super to a normal liquids.
The cutoff makes the dissipation weak. This accords with
the fact that the
nuclear dissipation plays less or no significant role in nuclear
fission at low temperatures\cite{N89}.
The pairing gap gets smaller as the temperature increases.
We suggested that the decay rate suddenly decreases at the
critical temperature, where the pairing gap disappears.
This could be related to the sudden decay of superdeformed band
at some critical angular momentum\cite{SVDB93}.

In this paper, we assumed the standard value for the pairing gap
parameter. The critical temperature was then found to be much lower than
the threshold temperature for the dissipative fission discussed in
ref.\cite{TB93}.
The non-monotonic behaviour of the decay rate
shown in Fig.3 in this paper might therefore indicate
the existence of the second critical temperature other than
the threshold temperature discussed in ref.\cite{TB93}.
The time dependent Hartree Fock (TDHF)
calculations for the induced fission of $^{236}$U
with the constraint of axial symmetry
uses, however, fairly large values of the effective pairing gap
which are about 2.5 to 7.5 times larger
than the usual values\cite{NKMNS78}.
The authors in ref.\cite{NKMNS78} claim that
these large values reflect the effect of the breaking of the axial symmetry.
If we use such large effective pairing gaps in our calculations,
the critical temperature, where the sudden decrease
of the fission rate due to the disappearance of the pairing gap
occurs, nearly coincides with the threshold temperature
found in ref.\cite{TB93}.
In order to draw a definite conclusion to this problem
more detailed studies of the coupling form factor
and of the temperature and the coordinate dependence of the
friction constant are required.
The work toward this direction is now in progress.

\bigskip

The authors thank Profs. H.A. Weidenm\"uller and D.M. Brink
for useful discussions.
The work of K.H. was supported by Research Fellowships
of the Japan Society for the Promotion of Science for
Young Scientists.
This work was supported by the Grant-in-Aid for General
Scientific Research,
Contract No.06640368, and the Grant-in-Aid for Scientific
Research on Priority Areas, Contract No.05243102,
from the Japanese Ministry of Education, Science and Culture.

\newpage

\begin{center}
{\bf Figure Captions}
\end{center}

\noindent
{\bf Fig.1:}
The cutoff frequency dependence of the cross over temperature $T_c$
between the quantum and the thermal regimes.
The solid line was obtained by numerically solving eq.(15).
The dashed and the dotted lines are the crossover temperature
in the absence of environments and that in the system with
Ohmic dissipation without cutoff, respectively.

\noindent
{\bf Fig.2:}
Quantum correction factor as a function of temperature.
The dashed and the dotted lines are the quantum correction factor
in the absence of environment and that in the system with
Ohmic dissipation without cutoff, respectively.
The solid line is the quantum correction factor when
a lower cutoff frequency has been introduced throuth the temperature
dependence of the pairing gap.
The left and the right arrows are the crossover temperature from a
quantal to a thermal decay, and the critical temperature for the
super to normal phase transition, respectively.

\noindent
{\bf Fig.3:}
Decay rate as a function of temperature.
The dashed and the dotted lines are the decay rate in the absence
of environment and in the Kramers limit, where there is no cutoff,
respectively. The solid line takes the effects of cutoff into account.


\newpage

\begin{thebibliography}{99}
\addtolength{\baselineskip}{-2mm}

\bibitem{BSAA93}D. Boilley, E. Suraud, Y. Abe, and S. Ayik,
Nucl. Phys. {\bf A556}, 67(1993).

\bibitem{WAC93}T. Wada, Y. Abe, and N. Carjan, Phys. Rev. Lett.
{\bf 70}, 3538(1993).

\bibitem{HHR89}D.J. Hinde, D. Hilscher, and H. Rossner, Nucl. Phys.
{\bf A502}, 497c(1989).

\bibitem{TB93}M. Thoennessen and G.F. Bertsch, Phys. Rev. Lett.
{\bf 71}, 4303(1993).

\bibitem{V94}R. Vandenbosch, Phys. Rev. {\bf C50}, 2618(1994).

\bibitem{F94}D. Fabris {\it et al.}, Phys. Rev. Lett. {\bf 73}, 2676
(1994).

\bibitem{FGM93}P. Fr\"obrich, I.I. Gontchar, and N.D. Mavlitov,
Nucl. Phys. {\bf A556}, 281(1993).

\bibitem{DW87}N.R. Dagdeviren and H.A. Weidenm\"uller, Phys. Lett.
{\bf B186}, 267(1987).

\bibitem{BBBV90}F. Barranco, G.F. Bertsch, R.A. Broglia, and
E. Vigezzi, Nucl. Phys. {\bf A512}, 253(1990).

\bibitem{LCBBV89}P. Lotti, F. Cazzola, P.F. Bortignon, R.A. Broglia, and
A. Vitturi, Phys. Rev. {\bf C40}, 1791(1989).

\bibitem{ACR87}F. Alassia, O. Civitarese, amd M. Reboiro, Phys. Rev.
{\bf C35}, 812(1987).

\bibitem{CDP83}O. Civitarese, G.G. Dussel, and R.P.J. Perazzo,
Nucl. Phys. {\bf A404}, 15(1983).

\bibitem{IG79}A. Iwamoto and W. Greiner, Z. Phys. {\bf A292}, 301(1979).

\bibitem{L67}J.S. Langer, Ann. Phys. (N.Y.) {\bf 41}, 108(1967);
G. Callan and S. Coleman, Phys. Rev. {\bf D16}, 1762(1977);
I.K. Affleck, Phys. Rev. Lett. {\bf 46}, 388(1981).

\bibitem{GOW87}H. Grabert, P. Olschowski and U. Weiss, Phys. Rev. {\bf B36},
1931(1987).

\bibitem {TA90} N. Takigawa and M. Abe, Phys. Rev. {\bf C41}, 2451(1990).

\bibitem{FT92}P. Fr\"obrich and G.-R. Tillack, Nucl. Phys. {\bf A540},
353(1992).

\bibitem{BZW94}J.-D. Bao, Y.-Z. Zhuo and X.-Z. Wu, Phys. Lett. {\bf B327},
1(1994); Z. Phys. {\bf A347}, 217(1994).

\bibitem{HTB90}P. H\"anggi, P. Talkner,and M. Borkovec, Rev. Mod. Phys.
{\bf 62}, 251(1990).

\bibitem{CL81}A.O. Caldeira and A.J. Leggett, Phys.Rev.Lett.
{\bf 46}, 211(1981).

\bibitem{THAB94}N. Takigawa, K. Hagino, M. Abe and A.B. Balantekin,
Phys.Rev.{\bf C49}, 2630(1994).

\bibitem{THA95}N. Takigawa, K. Hagino, and M. Abe,
Phys. Rev. {\bf C51}, 187(1995).

\bibitem{GSI88}H. Grabert, P. Schramm, and G.-L. Ingold,
Phys. Rep. {\bf 168}, 115(1988).

\bibitem{GJW83}P. Grang\'e, L. Jun-Qing, and H.A. Weidenm\"uller,
Phys. Rev. {\bf C27}, 2063(1983).

\bibitem{YAF93}H. Yoshioka, K. Awaka, and H. Fukuyama, (unpublished).

\bibitem{N89}J.R. Nix, Nucl. Phys. {\bf A502}, 609c(1989).

\bibitem{SVDB93}Y.R. Shimizu, E. Vigezzi, T. D{\o}ssing, amd R.A. Broglia,
Nucl. Phys. {\bf A557}, 99c(1993).

\bibitem{NKMNS78}J.W. Negele, S.E. Koonin, P. M\"oller, J.R. Nix,
and A.J. Sierk, Phys. Rev. {\bf C17}, 1098(1978).


\end{thebibliography}
\end{document}